\documentclass[aps,prb,twocolumn,superscriptaddress,longbibliography]{revtex4-2}
\usepackage{amsmath,amssymb}
\usepackage[pdftex]{hyperref,graphicx}
\hypersetup{colorlinks = true, urlcolor = blue, linkcolor = blue, citecolor = blue}
\usepackage{physics}
\usepackage{xcolor}
\usepackage{bm}
\newcommand{\anno}[1]{}

\begin{document}
\title{Interaction range and temperature dependence of symmetry breaking in strongly correlated two-dimensional moir\'e transition metal dichalcogenide bilayers}
\author{Haining Pan}
\affiliation{Condensed Matter Theory Center and Joint Quantum Institute, Department of Physics, University of Maryland, College Park, Maryland 20742, USA}
\author{Sankar Das Sarma}
\affiliation{Condensed Matter Theory Center and Joint Quantum Institute, Department of Physics, University of Maryland, College Park, Maryland 20742, USA}

\begin{abstract}
    We theoretically consider two-dimensional moir\'e transition metal dichalcogenide (TMD) bilayers, which are strongly correlated in the sense that the on-site Coulomb interaction is comparable to or larger than the hopping kinetic energy between the moir\'e lattice sites.  The system accommodates many symmetry-broken ground states both in charge and isospin sectors at various commensurate rational fillings such as 1/2,  1/3, 1/4, 2/3, etc. We investigate two complementary important aspects of the dependence of the symmetry breaking on (1) the range of the electron-electron interaction, which can in principle be experimentally controlled by the nearby gates and the dielectric environment, and (2) temperature, which could thermally suppress the symmetry breaking above a critical temperature.  Experimental implications of the theory are discussed.
\end{abstract}

\maketitle
\textit{Introduction.} Two-dimensional (2D) moir\'e transition metal dichalcogenide (TMD) bilayers have recently emerged as a tunable interacting lattice system to study strong correlation effects in semiconductor layers~\cite{regan2020mott,wang2020correlated,xu2020correlated,zhang2020flat,zhou2021bilayer,tang2020simulation,liu2021excitonic,li2021imaging,li2021continuous,li2021quantum,huang2021correlated,jin2021stripe,ghiotto2021quantum,li2021chargeorderenhanced}.\anno{[cite all good relevant experimental papers]} It was pointed out that these systems are ideal analog solid state emulators of the 2D Hubbard model in the strongly correlated regime with $U/t \sim O(1)$, where $U$ and $t$ are the usual effective electron interaction and noninteracting kinetic energy scales, respectively, as appearing in the Hubbard model~\cite{wu2018hubbard,hu2021competinga,morales-duran2021nonlocal,morales-duran2021metalinsulator,pan2020band,pan2020quantum,pan2021interactiondriven,zang2021hartreefock}. \anno{[cite the relevant papers from Texas and our three papers and the paper from Flatiron and any other relevant and correct theory papers]}  Various charge and isospin symmetry broken phases have been theoretically predicted and experimentally observed in both TMD homo- and hetero bilayers (sometimes in the presence of an applied perpendicular electric field $V_z$, which modifies the noninteracting band structure, further enhancing the dimensionless interaction strength), and there is a general agreement between theory and experiment, although theory seems to predict richer exotic possibilities (e.g., spin liquids, superconductors) than have been experimentally validated so far.  But the most significant theoretical predictions\anno{[cite our work here]} of correlated insulators at half filling and the existence of generalized Wigner crystal type charge density waves (CDWs) have been directly verified experimentally~\cite{li2021imaging}.\anno{[cite the very recent Nature paper from Berkeley]} Also, the predicted doping- and field-tuned insulator-to-metal transitions have also been recently observed~\cite{li2021continuous,ghiotto2021quantum,pan2021interactiondriven} \anno{[cite Cornell and Columbia Nature papers, and our PRL here]} at half filling, but the role of disorder is likely to be important in such transitions~\cite{ahn2021disorder}. \anno{[cite Ahn-SDS preprint]}  The subject is undoubtedly among the most active topics in all of condensed matter and strong correlation physics, and as such, many important questions remain unaddressed.

In the current Letter, we investigate two such open questions, finding some surprising results.  This Letter consists of two complementary issues, both with immediate and important experimental and theoretical significance.  The first topic is the role of the interaction range on the symmetry-breaking phenomenon.  In particular, the bare inter particle (electron or hole) interaction is by definition the long-ranged Coulomb interaction screened by the environment, and thus a minimal Hubbard model with just a moir\'e on-site interaction with a single $U$ is inappropriate for TMDs.  We systematically investigate the effect of longer-range coupling between far away moir\'e sites on the ground state symmetry breaking, finding that symmetry breaking may actually sometimes be suppressed by having distant neighbor Coulomb coupling, a result which sounds counterintuitive, but is nevertheless correct.  The second topic we study is the thermal suppression of the symmetry-broken phase with the charge gap vanishing at a critical temperature, as has recently been reported experimentally~\cite{ghiotto2021quantum}. \anno{[cite the recent Columbia Nature]}

\textit{Theory.} Our model is a generalized Hubbard model on a triangular moir\'e lattice realized in a twisted homobilayer WSe$_2$ as per
\begin{equation}\label{eq:H}
    \begin{split}
        &H=\sum_{s}\sum_{i,j}^{} t_{s}\left(\bm{R}_i-\bm{R}_j\right) c_{i,s}^\dagger c_{j,s}\\
        &+\frac{1}{2}\sum_{s,s'}\sum_{i,j}U(\bm{R}_i-\bm{R}_j) c_{i,s}^\dagger c_{j,s'}^\dagger c_{j,s'} c_{i,s},
    \end{split}
\end{equation}
where $t_{s}\left(\bm{R}_i-\bm{R}_j\right)$ is the hopping between the moir\'e lattice site $i$ and $j$, fitted from the first moir\'e valence band of twisted WSe$_2$ at $\pm K$ valleys in the noninteracting picture, and $s=\uparrow$ and $ \downarrow$ are coupled with the $+K$ and $-K$ valleys~\cite{pan2020band}. $U(\bm{R}_i-\bm{R}_j)$ describes the Coulomb repulsion between site $i$ and $j$. We control the range of the Coulomb interactions by including a finite number of neighbors coupled in the theory. These neighbors are denoted by hexagonal ``shells''  (e.g., on-site interaction $U_0$ is 0-shell, 1-shell means six nearest-neighbor interactions $U_1$ plus $U_0$) in the following results. By comparing the results of short-range (few shells) and long-range (many shells) interactions, we investigate the role of the interaction range.

To find the ground states of Eq.~\eqref{eq:H}, which contains quartic terms, at the mean-field level, we use the Hartree-Fock approximation to expand the quartic interaction [the second term in Eq.~\eqref{eq:H}] into quadratic terms as $\sum_{s,s'}\sum_{i,j} U(\bm{R}_i-\bm{R}_j)\left( \expval{c_{i,s}^\dagger c_{i,s}} c_{j,s'}^\dagger c_{j,s'}-\expval{c_{i,s}^\dagger c_{j,s'}} c_{j,s'}^\dagger c_{i,s} \right)$, and diagonalize the quadratic Hamiltonian in the momentum space to find the self-consistent solution starting from a given initial ansatz. In the end, we obtain the energetically favorable ground state by comparing the energies of various candidate phases (Mott insulator, correlated insulator, normal metal, etc). We refer the reader to Ref.~\onlinecite{pan2020quantum} for details.

To address the second question, we consider the finite-temperature thermal state at a fixed filling factor.  Here, the temperature enters through the correlator $\expval{c_{i,s}^\dagger c_{j,s'}}$, which is averaged with respect to the occupation number of each state with energy $E$, where the occupation number follows the Fermi-Dirac distribution $f_T(E)=\left\{ 1+\exp[(E-\mu)/T] \right\}^{-1}$ at temperature $T$. At finite temperature, we define the filling factor by the average number of holes per site as $\nu=\frac{1}{N}\sum_{i,s} \expval{c_{i,s}^\dagger c_{i,s}}$, where $N$ is the total number of moir\'e sites in the system, and $\nu=0$ corresponds to the charge neutrality point.
Thus, the chemical potential $\mu$ must be adjusted in each iteration to ensure the invariance of the filling factor $\nu$. 
To study the metal-insulator transition (MIT) driven by temperature $T$, we use the charge gap $E_G$, defined by the extra energy of adding one hole to the system, as the order parameter of the insulating phase ($E_G>0$) and metallic phase ($E_G=0$), following the experiment~\cite{ghiotto2021quantum}. 

The phase transitions in moir\'e TMD are extremely rich as they can be driven by many different parameters, such as the interaction range, $V_z$, $ \epsilon$, $\theta$, $ \nu$, and $T$. Here, the interaction range determines the nature of the interaction in Eq.~\eqref{eq:H}: whether it is a short-range screened Coulomb interaction or a long-range pure Coulomb interaction. The perpendicular electric field $V_z$ effectively generates a Dzyaloshinskii-Moriya interaction at $\nu=1$~\cite{pan2020band}, which alters the single-particle band structure, and consequently changes the magnitude, adding a complex phase to the hopping parameter $t$ in Eq.~\eqref{eq:H} (refer to Fig. 10 in Ref.~\onlinecite{pan2020band} for the magnitude and phase of $t$ as a function of $V_z$). The dielectric constant $\epsilon$ and the twist angle $\theta$ control the strength of the Coulomb interaction and hopping, respectively, which can induce the transition from ferromagnetic to antiferromagnetic order by changing the exchange energy $t^2/U$.  The temperature $T$ can thermally suppress the insulating energy gap to induce a MIT. 

\begin{figure*}[ht]
    \centering 
    \includegraphics[width=5in]{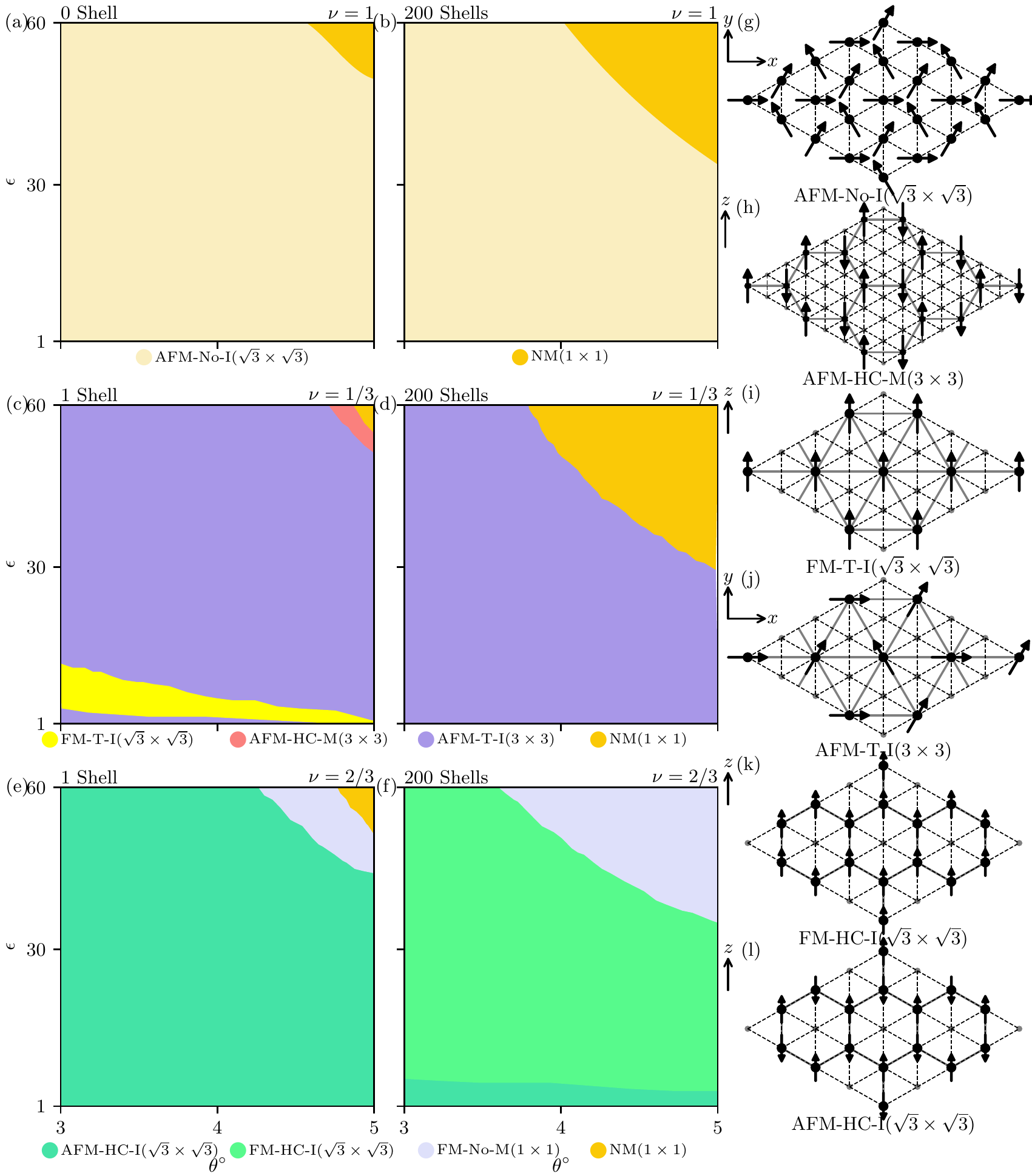}
    \caption{Phase diagrams as a function of the twist angle $\theta$ and dielectric constant $\epsilon$ at (a) and (b) $\nu=1$; (c) and (d) $\nu=1/3$; (e) and (f) $\nu=2/3$. The left column is for the short-range interaction while the right column is for the long-range interaction. Here, ``T'' means the triangular lattice, ``HC'' means the honeycomb lattice, ``No'' means no CDW, ``NM'' means spinless normal metal, ``M'' means the phase is a metal, and ``I'' means the phase is an insulator. The numbers in the parentheses indicate the extended periods of each phase which are shown in (g)-(l).
    }
    \label{fig:1}
\end{figure*}

\textit{Results and discussion.} We first set the electric field $V_z$ and temperature $T$ to zero to investigate the interaction range.
Here, the short-range interaction means only on-site (for $\nu=1$) or on-site plus nearest neighbor (for $\nu<1$). The long-range interaction means distant neighbors (up to 200-shell) are coupled by $U$ in the theory.
We use the strict Coulomb interaction, $V(r) = e^2/(\epsilon r)$, where $r$ is the Wannier-weighted inter particle separation as appropriate for the TMD moir\'e system and $\epsilon$ is the effective background dielectric constant including the substrates, and calculate the mean-field Hartree-Fock quantum phase diagram numerically self-consistently starting from the classical ansatz for the charge density wave states as in Refs.~\onlinecite{pan2020band,pan2020quantum}\anno{ [ cite earlier works where this is clearly explained]}.  
We show in Fig.~\ref{fig:1} the resultant $\epsilon-\theta$ phase diagrams for filling $\nu$=1, 1/3, 2/3 where $\nu$=1 implies one hole per moir\'e unit cell (i.e., half filled in the usual Mott-Hubbard terminology).  For each filling, we show side by side the results for just the short range (left panels) and long range (right panels), where the results have basically converged to the infinite shell limit.  
The results presented in Fig.~\ref{fig:1} involve first carrying out the moir\'e band structure calculation for each value of $\epsilon$ and $\theta$, and then taking out the ground state of the moir\'e band structure to calculate the hopping term $t$ and Coulomb interaction term $U$ to construct the effective low-energy lattice model--- the (extended) Hubbard model. Next, we solve the many-body Hamiltonian using the Hartree-Fock approximation by first contriving an ansatz wave function and diagonalizing the Hamiltonian iteratively to obtain a self-consistent solution. By comparing the energies of different competing states, we determine the ground state at a specific parameter. In the end, we change multiple parameters to obtain the full phase diagram.
We refer the reader to Refs.~\onlinecite{pan2020quantum,pan2020band} for details of these phases.

The results in Fig.~\ref{fig:1} are surprising to say the least: For example, in Figs.~\ref{fig:1}(a) and~\ref{fig:1}(b) at $\nu=1$, the system undergoes a MIT from an insulator with $120^\circ$ antiferromagnetic N\'eel state [denoted by ``AFM-No-I($\sqrt{3}\times \sqrt{3}$)'' in Fig.~\ref{fig:1}(g)] to normal gapless metallic phase as $\theta$ increases or $\epsilon$ increases. 
The normal gapless metallic phase (i.e., the non-symmetry-broken phase corresponding to the noninteracting band structure prediction at half filling) is preferred for large $\epsilon$ and $\theta$ by the full long-range interaction [Fig.~\ref{fig:1}(b)] over the corresponding 0-shell short-range interaction calculation [Fig.~\ref{fig:1}(a)], although the 200-shell calculation includes $U_0$ and $U_1$ and a large number of additional interaction terms all the way to 200-shell.  
A similar situation also arises for $\nu$=1/3 (i.e., 1/6 filling) in Figs.~\ref{fig:1}(c) and~\ref{fig:1}(d), where the full long-range 200-shell calculation [Fig.~\ref{fig:1}(d)] has a much larger regime of the non-symmetry-broken normal gapless metallic phase than the calculation including only $U_0$ and $U_1$ [Fig.~\ref{fig:1}(c)].
By contrast, the $\nu$=2/3 (i.e., 1/3 filling) manifests more complicated behavior for the full 200-shell long-range theory [Fig.~\ref{fig:1}(f)] than the short-range $U_0 + U_1$ model [Fig.~\ref{fig:1}(e)], with the normal metallic phase (upper right-hand corner) being a symmetry-broken ferromagnetic metallic phase rather than a normal nonmagnetic metal as for the $U_0 + U_1$ short-range model.  In addition, the dominant state ``AFM-HC-I($\sqrt{3}\times\sqrt{3}$)'' [ Fig.~\ref{fig:1}(l)] with antiferromagnetic order switches to ``FM-HC-I($\sqrt{3}\times\sqrt{3}$)'' [Fig.~\ref{fig:1}(k)] with ferromagnetic order. The qualitative difference between the predicted phase diagrams in the short-range and long-range interaction models can be understood as arising from a subtle interplay between hopping, Pauli principle, and interaction range, which has no generic easy and intuitive explanation.  It is easy to see in a 1D lattice that having just the on-site interaction and simply adding one next-nearest-neighbor interaction term drastically alters the configuration-dependent local ground state energetics, and the situation is obviously extremely complex on a 2D triangular lattice with multiple hopping and interaction terms with no easy and obvious explanations for the details of the phase diagram--- one must do the calculation to obtain the correct ground states, and as is obvious from Fig.~\ref{fig:1}, making the standard on-site Hubbard model approximation uncritically may be misleading.

Our results in Fig.~\ref{fig:1} are just not surprising, as they have clear experimental implications since, in principle, the interaction range can be controlled by introducing metallic gates close to the TMD bilayer.  In addition, samples with varying $\theta$ and a varying background dielectric constant (by using different substrates) can be fabricated to directly verify that the resultant quantum phase diagram depends quite sensitively on the interaction range.  It is actually possible that our predictions of Fig.~\ref{fig:1} have already been verified experimentally as samples from different laboratories do not always reflect the same symmetry-broken phases, but without a systematic experimental study, one cannot be sure that the striking physics predicted in Fig.~\ref{fig:1} has been observed.

\begin{figure}[ht]
    \centering
    \includegraphics[width=3.4in]{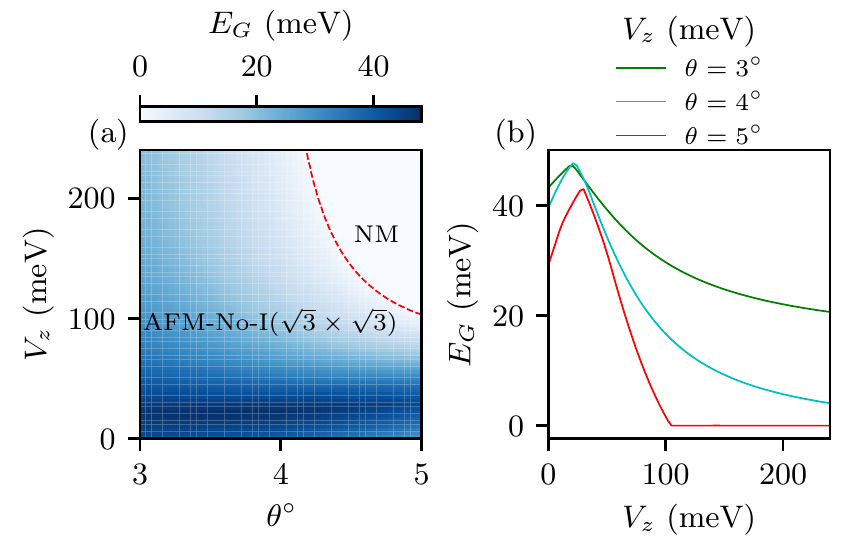}
    \caption{(a) Charge gap $E_G$ as a function of the twist angle $\theta$ and the electric field $V_z$ at $\epsilon=20$ and $\nu=1$ . The red dashed line is the phase boundary between ``AFM-No-I($\sqrt{3}\times\sqrt{3}$ )'' and normal metal. (b) Line cuts of the charge gap $E_G$ as a function of $V_z$ for $\theta=3^\circ$ (green), $\theta=4^\circ$ (blue), and $\theta=5^\circ$ (red).}
    \label{fig:2}
\end{figure}

\begin{figure}[ht]
    \centering
    \includegraphics[width=3.4in]{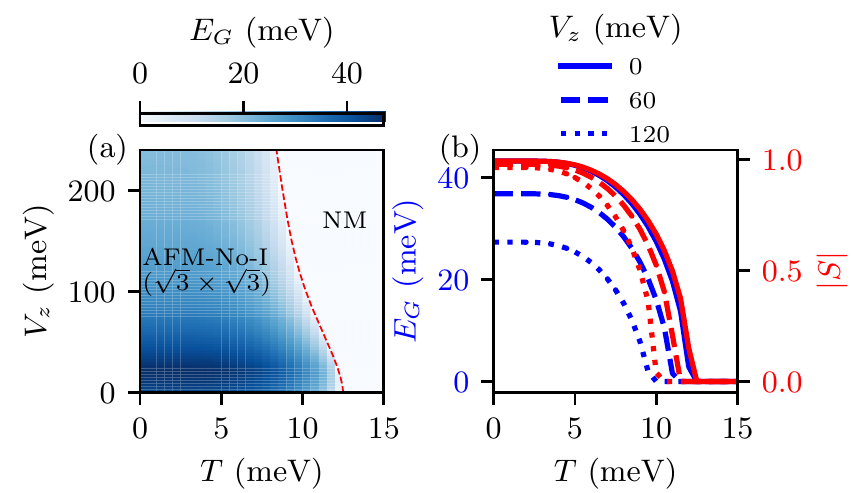}
    \caption{(a) Charge gap as a function of the electric field $V_z$ and the temperature $T$ at $\theta=3^\circ$, $\nu=1$ and $\epsilon=20$. The red dashed line indicates the $T_c$, which is the phase boundary between ``AFM-No-I($\sqrt{3}\times\sqrt{3}$ )'' and normal metal. (b) Linecuts of the charge gap $E_G$ (blue, left axis) and the spin magnitude $\abs{S}$ (red, right axis) as a function of the temperature $T$ for $V_z=0$ (solid line), $V_z=60$ meV (dashed line), and $V_z=120$ meV (dotted line). }
    \label{fig:3}
\end{figure}

In discussing the thermal MIT and a possible critical temperature at $\nu$=1, as observed recently~\cite{ghiotto2021quantum}\anno{ [Columbia Nature 2021]},  we focus on $\nu$=1 and consider only the 0-shell theory, which agrees with the full long-range results up to large $\epsilon$  ($<$30) according to Figs.~\ref{fig:1}(a) and~\ref{fig:1}(b).  It is important to realize that the charge gap can be closed even at $T$=0 causing a MIT by tuning $\epsilon$  (controlling interaction), $\theta$  (controlling hopping), or an applied electric field $V_z$ which also modifies the single-particle band structure affecting the effective correlation parameter $U/t$. In Fig.~\ref{fig:2}(a), we first provide our calculated phase diagram as a function of $\theta$ and $V_z$ at $T$=0 using the same self-consistent mean-field Hartree-Fock theory as used for Fig.~\ref{fig:1}~\cite{pan2020band}.\anno{[cite our PRR]}  We fix $\epsilon=20$ for these calculations, which is our best estimate for the screening environment for Ref.~\onlinecite{ghiotto2021quantum}\anno{[Columbia]} including both the substrate and the gate-induced screening. (Results for other $\epsilon$ values are similar with the gap energies and phase boundary being quantitatively different.)
The phase boundary of the metal-insulator transition is indicated by the red dashed line, where the metallic phase prefers large twist angles $\theta$ and large $V_z$.  We see from Fig.~\ref{fig:2}(b) that large $V_z$ by itself can close the charge gap of the antiferromagnetic insulating phase at $V_z\sim 100$ meV for $\theta =5^\circ$ (red line), but not for a smaller twist angle [e.g., $\theta=3^\circ$ (green line)]. So, our first prediction, consistent with recent experiments, is that finite $V_z$ would cause an insulator-to-metal Mott transition at half filling for larger twist angles.

Since the metal-insulator transition cannot be induced by $V_z$ for the small angle ($\theta\sim 3^\circ$), an interesting question is whether the finite temperature can induce a MIT by thermally suppressing the charge gap. In Fig.~\ref{fig:3}, we show our calculated mean-field thermal $T-V_z$ phase diagram for $\nu$=1, $\theta=3^\circ$, and $\epsilon$=20.  We emphasize that $V_z$  by itself cannot cause a MIT at $\theta=3^\circ$, with the ground state being the ``AFM-No-I($\sqrt{3}\times\sqrt{3}$)'' with a charge gap for all $V_z$, but finite temperature destroys the symmetry-broken long-range order through thermal fluctuations with a critical $T_c\sim$10 meV (which is an upper bound since our mean-field treatment ignores fluctuations).  We note two experimentally verifiable important features of Fig.~\ref{fig:3}: (1) $T_c$ [the red dashed line in Fig.~\ref{fig:3}(a)] does not depend strongly on $V_z$, with $T_c $ varying by less than 25\% for very large changes (from 0 to 240 meV) in $V_z$; (2) both the charge gap $E_G$ [blue lines in Fig.~\ref{fig:3}(b)] and the spin texture, denoted by its magnitude $\abs{S}$ [red lines in Fig.~\ref{fig:3}(b)], drop to zero at the same $T_c$ for all $V_z$ (solid, dashed, and dotted lines corresponding to the results at $V_z$=0, 60, and 120 meV, respectively). The suppression of the charge gap $E_G$ and spin magnitude $\abs{S}$ as a function of temperature manifests a typical feature of self-consistent solutions from the mean-field Hamiltonian.  The phase for $T>T_c$ is thus a gapless paramagnetic normal metal with no long-range spin ordering. This is also a verifiable prediction.  The fact that the thermal suppression of the antiferromagnetic insulator phase [``AFM-No-I($\sqrt{3}\times\sqrt{3}$)''] is not sensitively dependent on $V_z$ [rather obvious in Fig.~\ref{fig:3}(a)]  is another verifiable prediction.

\textit{Conclusion.} We study theoretically the quantum phase diagrams of moir\'e TMD systems, addressing the complementary physics of the effects of interaction range, applied electric field, and temperature on the symmetry-breaking properties.  We find that increasing the interaction range has an apparent surprising effect of actually suppressing the symmetry breaking in some situations.  We calculate the finite-temperature TMD thermal phase diagram, finding that both the charge gap and the spin texture disappear at half filling above a critical temperature simultaneously, leading to a thermal phase transition from an antiferromagnetic gapped insulator to a gapless paramagnetic metal.  Our predictions are directly experimentally verifiable, and both the temperature-induced and the electric field-induced insulator-to-metal transition at half filling have already been observed~\cite{li2021continuous,ghiotto2021quantum}.

This work is supported by the Laboratory for Physical Sciences. We also acknowledge the support of the University of Maryland High-Performance Computing Cluster (HPCC).

\bibliography{Paper_TMD.bib}

\end{document}